\documentstyle[aps,12pt,epsfig]{revtex}
\def\laq{\ \raise 0.4ex\hbox{$<$}\kern -0.8em\lower 0.62
ex\hbox{$\sim$}\ }
\def\gaq{\ \raise 0.4ex\hbox{$>$}\kern -0.7em\lower 0.62
ex\hbox{$\sim$}\ }

\def\NPB{{\em Nucl. Phys.} B}
\def\PLB{{\em Phys. Lett.}  B}

\def\PRD{{\em Phys. Rev.} D}

\begin{document}


\title{Cold and Hot Dark Matter from a Single Nonthermal Relic}

\author{Ram Brustein} 
\address{Department of Physics,
Ben-Gurion University,
Beer-Sheva 84105, Israel\\
email: ramyb@bgumail.bgu.ac.il}

\author{Merav Hadad} 
\address{School of Physics and Astronomy,
Beverly and Raymond Sackler Faculty of Exact Sciences,\\
Tel Aviv University, Tel Aviv 69978, Israel\\
email: meravv@post.tau.ac.il}

\maketitle
\begin{abstract}
The origin of dark matter in the universe may be 
scalar particles produced by amplification of quantum fluctuations during a
period of dilaton-driven inflation. We show, for the first time, 
that a single species of particles, 
depending on its mass and interactions,  can be a source of both cold and hot
dark matter simultaneously. Detection of such weakly interacting
particles with masses below a fraction of an eV presents a new challenge for 
dark matter searches.

 \end{abstract}


It is widely accepted that a substantial fraction of the total 
energy density in the universe is in the form  of dark matter (DM). 
The composition  and amount of DM is not known, but many hypothetical 
particles  were proposed as candidates:
WIMPs, axions, LSP's, massive neutrinos and more \cite{dmrev,eu}. 
DM is classified
into two types according to the velocities of particles at the beginning of
the epoch of structure formation when the temperature of the universe was about
1 eV, cold dark matter (CDM)  if the particles are non-relativistic and hot 
dark matter (HDM) if they are relativistic. 
The prevailing wisdom is that CDM and HDM originate from different 
species of particles, for example, axions
which weigh a fraction of an eV for CDM, and  a few eV
neutrinos for HDM. DM particles are traditionally 
assumed to have a thermal distribution of velocities (see, however, 
\cite{kolb}).
Since a  thermal distribution is sharply peaked around a single velocity,
if that velocity is relativistic, the amount of
non-relativistic particles is extremely small and vice versa.

We show, for the first time,  that weakly interacting nonthermal relics 
produced by amplification of quantum fluctuations during a period of 
dilaton-driven inflation can have an energy spectrum with two peaks,  
such that at the beginning of the epoch of structure formation some 
fraction of the particles are relativistic and some fraction are 
non-relativistic. The ratio and magnitude of energy densities in 
relativistic and nonrelativistic particles are determined by the 
mass of the particles and by the cosmological parameters of the model. 
The basic physics behind the appearance of a twin peak spectrum is the 
existence of  two scales in the problem: the Planck scale, redshifted, 
and the mass.
Our models suggest the possibility that DM in the universe might 
be composed of weakly interacting nonthermal relics and that it might 
be a mixture of CDM and HDM if the mass of the particles is lighter than a 
fraction of an eV. It is argued \cite{primack} that CDM+HDM models
could be just what is needed to explain the data on CMB anisotropies 
\cite{pdg}. Our models can provide the desired composition from a 
single source, as well  as predict other observable consequences.

We consider particle production in models of string cosmology which
realize the pre-big-bang scenario \cite{sfd,pbb}. In this scenario the evolution
of the universe starts from a state of very small curvature and coupling and
then undergoes a long phase of dilaton-driven kinetic inflation  and at some
later time joins smoothly standard  radiation dominated cosmological evolution,
thus giving rise to a singularity free inflationary cosmology.
Particles are produced during the
period of dilaton-driven inflation by the standard mechanism of amplification of
quantum fluctuations \cite{mukh}. 

In the simplified model of background evolution we adopt, the evolution of the
universe is divided into four distinct  phases  with specific (conformal)
time dependence of the scale factor of the universe $a(\eta)$ and the dilaton
$\phi(\eta)$.  The first phase is a long 
dilaton-driven inflationary phase, the second phase is a high curvature string
phase of otherwise unknown properties, followed by ordinary
Friedman-Robertson-Walker  (FRW) radiation dominated (RD) evolution and then a
standard FRW matter dominated (MD) evolution.  We assume throughout an isotropic
and homogeneous four dimensional flat universe, described by a FRW metric. All
other scalar fields  are assumed to have a trivial vacuum expectation value 
during the inflationary phase. 

During the dilaton-driven inflationary phase $\eta<\eta_s$, 
both scale factor  and coupling
$e^\phi$ are growing  as powers $a(\eta )=a_s\left( \frac \eta {\eta _s}\right)
^\alpha$ and $ e^{\phi (\eta )}=e^{\phi _s}\left( \frac \eta {\eta _s}\right)
^\beta$, where $\alpha$ and $\beta$ are negative. The dilaton-driven phase is
expected to last until curvatures reach the string scale  and the background
solution starts to deviate substantially from the lowest order solution. For
ideas about how this may come about see \cite{exit}. The string phase lasts 
while  $\eta_s<\eta <\eta_1$. We assume that curvature stays high during the string
phase. As in \cite{peak}, we assume that  the string phase ends when curvature
reaches the string scale $M_s$.  We parametrize our
ignorance about the string phase background, as in \cite{bggv}, by the ratios of
the scale factor and the string coupling $g(\eta)=e^{\phi(\eta)/2}$,  
at the beginning
and end of the string phase $z_S=a_1/a_S$ and $g_1/g_S$, where
$g_1=e^{\phi(\eta_1)/2}$ and  $g_S=e^{\phi(\eta_S)/2}$, where $a_S=a(\eta_s)$ and
$\phi_S=\phi(\eta_s)$.    We take the parameters to be in a range we consider
reasonable. For example, $z_S$ could be in the range $1<z_S<e^{45}\sim 10^{20}$,
to allow a large part of the observed universe to originate in the
dilaton-driven phase, and $g_1/g_S>1$, assuming that the coupling continues to
increase during the string phase and $10^{-3}\laq g_1\laq 10^{-1}$ to agree with the
expected range of string mass (see e.g. \cite{peak}). Some other useful
quantities that we will need are $\omega_1$, the frequency today, corresponding
to the end of the string phase, estimated in \cite{peak} to be $\omega_1\sim
10^{10}Hz$, and the frequency $\omega_S= \omega_1/z_S$, the frequency today
corresponding to the end of the dilaton-driven phase.
In the RD phase and MD phase are assumed to follow the string 
phase, the dilaton is taken to be strictly constant, frozen at its value
today.

We have computed the spectrum of produced particles for the models 
 described previously \cite{bh1}. We have solved a linear 
perturbation equation, 
$
\chi _k^{\prime \prime }+\left( k^2+M^2a^2-\frac{s^{\prime \prime }}s\right)
\chi _k=0
$
where 
$
s(\eta )\equiv a(\eta )^me^{l\phi (\eta )/2}=a_s^me^{l\phi _s/2}\left( \frac
\eta {\eta _s}\right) ^{1/2-n_s},
$
imposing initial conditions corresponding to normalized vacuum fluctuations.
The parameter $m$ depends on the spin of the particle and $l$ depends on
its coupling to the dilaton.
The perturbation first ``exit the horizon" when $k \eta\sim 1$, 
when curvatures become larger than their wavelength, then
they are ``frozen" outside the horizon when $k\eta<1$, and 
then ``reenter the horizon" at $\eta_{re}$.
A duality symmetry \cite{sduality} exchanging the perturbation and its 
conjugate momentum can be used to follow the evolution of the 
perturbation equations for times
at which the background evolution is not known precisely.  
To read off the spectrum for a more general case of background
evolution that we consider here,  the only required
substitution in the results of  \cite{bh1} is  
$
n_s=\frac 12-\alpha m-\frac \beta 2l.
$ (see also \cite{bh2}).
Similar calculations have been performed by several groups  
and the results agree \cite{cllw,bmuv,giov98}, 
whenever a comparison was possible.

We will consider weakly interacting scalar particles, abundant in string
theory and supergravity. For scalar fields $m=1$, and we will consider 
for concreteness the following values for $l$,
$l=-1,0,1$ corresponding, respectively, to moduli (including the dilaton),
Ramond-Ramond axions, and Neveu-Schwartz axions. We will assume that 
the produced particles interact so weakly, that their interactions
and decay are not sufficient to alter the primordial spectrum substantially.
The particles we have in mind have typically gravitational strength 
interactions, which is definitely weak enough to satisfy our assumption, and
 masses below a fraction of an eV.

A typical spectrum of a light scalar may be divided, at a given time, 
into three frequency regions:
$i)$ The massless region, $\omega_S>\omega>M$. 
In this region particles are relativistic. 
$ii)$ The ``false'' massive region,  $M>\omega>\omega_m$, where
$\omega_m=\omega_1(M/M_s)^{1/2}$. 
In this region particles are nonrelativistic, 
but have reentered the horizon as relativistic modes.
$iii)$ The ``real'' massive region $\omega_m>\omega$. 
In this region particles are non-relativistic, 
and have reentered the horizon as non-relativistic modes. Note that 
physical frequencies redshift as the universe expands, and therefore 
boundaries of  regions change in time.

Different spectral shapes may result depending on parameters. 
For the interesting case, 
the spectrum  increases with $\omega$ in the massless region, 
 decreases in the ``false'' massive region and increases 
in the ``real" massive region. In this case, the energy density
in relativistic particles
$
\Omega _{REL}\simeq \frac{d\Omega }{d\ln w}\left( \omega=\omega_S\right) 
$, and that of nonrelativistic particles is given by
$
\Omega _{NR}\simeq \frac{d\Omega }{d\ln w}\left( \omega=\omega_m\right)
$.
For this case the spectrum takes the following approximate form at the 
beginning
of the epoch of structure formation $\eta_{eq}$,
\begin{equation}
 \frac{d\Omega }{d\ln \omega} = 
\cases{
 {\cal N}
g_1^2\left( \frac{g_S}{g_1}\right) ^{-2l}
\left( \frac {\omega}{\omega_S}\right) ^x  
& $\omega_S>\omega>M$,   \cr 
{\cal N}
 g_1^2\left( \frac{g_S}{g_1}\right) ^{-2l}\frac {M}{\omega_S}\left( \frac
{\omega}{\omega_S}\right) ^{x-1}  
&  $M>\omega>\omega_m$,   \cr 
{\cal N}
 g_1^2 \left( \frac{g_S}{g_1}\right) ^{-2l}
\frac {M}{\omega_1}\left( \frac {M}{M_s}\right) ^{-1/2} 
\left( \frac {\omega}{\omega_S}\right) ^x  
 & $\omega_m > \omega$,   }
\label{chspectrum}
\end{equation}
where 
$
x\equiv 2+2\alpha +l\beta 
$, and ${\cal N}$ is a numerical factor, estimated in \cite{bh1}, which
we will set to unity in what follows.

We have ignored, so far, particles produced during the string phase, since
that phase  is at the moment less well understood. If the spectrum decreases
there, then our approximations remain valid. If, however, 
the spectrum increases  there, 
then a good approximate relation to use would be
$
\Omega _{REL}\simeq \frac{d\Omega }{d\ln w}\left( \omega=\omega_1\right). 
$
Otherwise more parameters describing the string phase should be added. 
Since this is not relevant to our main point, we choose not to do so.

We impose constraints on the spectrum, and show that it 
is possible to satisfy all of  them by giving a specific example.
First we require that the desired spectral shape is obtained
\begin{equation}
 0<x<1.
\label{fstcon}
\end{equation}
Then we require that some relativistic DM particles and some nonrelativistic
 DM particles exist at $\eta_{eq}$,
\begin{equation}
 M<\omega_S(\eta_{eq}),\ \hspace{.5in}
\omega_m>\omega_{eq},
\label{scndcon}
\end{equation}
where $\omega_{eq}= \frac{1}{a^2(\eta_{eq})} \frac{d a}{d\eta}(\eta_{eq})$.
We also require that the energy density in produced particles is
the main source of energy density in the universe and that approximately
equal amounts exists (later we will see that this condition can be 
significantly relaxed).
\begin{equation}
\Omega _{massive}(t_{eq})\simeq 1, \ \hspace{.5in}
\Omega _{massless}(t_{eq}) \simeq 1.
\label{thrdcon}
\end{equation}
Assuming that the level of fluctuations at very large scales is
well below the CMB level (as we show later this must be the case),
we require
\begin{equation}
\frac{d\Omega}{d\ln \omega}\left( \omega=\omega_0\right) < 10^{-5}.
\label{frthcon}
\end{equation}

\begin{figure}
\begin{flushleft}
\psfig{figure=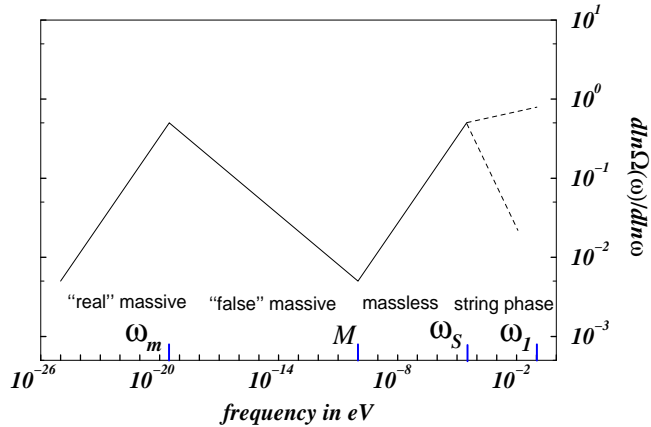,height=3.5in,width=2.5in}
\end{flushleft}
\caption{A typical twin peaked spectrum. The dashed lines represents
two possible string phase spectra.
\label{fig:TS} }
\end{figure}

A typical interesting spectrum is shown in Figure \ref{fig:TS},
\vspace{-1.2in}

The constraints on the spectral parameters can be translated into
constraints on cosmological parameters.
The range of powers $\alpha$  and $\beta$ allowed by condition (\ref{fstcon})
is summarized in Table \ref{tab:SI},

\vbox{
\begin{table}
\caption{Allowed range of cosmologies}
\begin{tabular}{ccc}
$l=-1$ & $l=0$ & $l=1$  \\
\tableline \noalign{\vskip2pt}
$\alpha <0$ & $-1<\alpha <-\frac{1}{2}$ & $\alpha <0$  \\
$\beta <0$  & $\beta <0$ & $-2<\beta <0$  \\
$\frac{\beta}{2}-1<\alpha <\frac{\beta}{2}-\frac{1}{2}$&  
& $-\!\frac{\beta}{2}\!-\!1<\alpha <-\!\frac{\beta}{2}\!-\frac{1}{2}$
\end{tabular}
\label{tab:SI}
\end{table}
}

\noindent
Condition (\ref{scndcon}) leads to
\begin{equation}
M<\omega_1(\eta_{eq}) z_S^{-1}, \hspace{.5in} 
M> M_s  \left(\frac{\omega_{eq}}{\omega_1}\right)^2.
\label{cp3}
\end{equation}
Condition (\ref{thrdcon}) leads to
\begin{equation}
 g_1^2\left( \frac{g_1}{g_S}\right)^{2l}\simeq 1,
\label{cp3a}
\end{equation}
and 
\begin{equation}
  g_1^2 \left( \frac{g_1}{g_S}\right) ^{2l}
\frac M{\omega_1}\left( \frac{\omega_m}{\omega_S}\right)^x
\left( \frac {M_s}{M}\right) ^{1/2}\simeq 1.
\label{cp3b}
\end{equation}
Since $\omega_m=\omega_1(M/M_s)^{1/2}$ and $\omega_S=\omega_1 z_S^{-1}$,
we obtain by substituting (\ref{cp3a}) into (\ref{cp3b})
$
\left( \frac {M}{M_s}\right)^{(1+x)/2}\frac{M_s}{\omega_1}z_S^x\simeq 1.
$
Condition (\ref{frthcon}) leads to 
$
g_1^2\left( \frac{g_1}{ g_S}\right) ^{2l}
\frac M{\omega_1}\left( \frac{\omega_0}{\omega_S}\right) ^x
\left( \frac {M_s}{M}\right) ^{1/2}<10^{-5},
$
which, using (\ref{cp3a}), leads to 
$
\left( \frac M{M_s}\right) ^{1/2}\frac{M_s}{%
\omega_1}\left( \frac{\omega_0}{\omega_1}\right) ^xz_S^x < 10^{-5}.
$
We therefore obtain the following two conditions, 
\begin{eqnarray}
M&>&10^{10/x}M_s\left( \frac{\omega_0}{\omega_1}\right) ^2,  \nonumber\\
z_S& < &10^{-5(1+x)/x^2}
\left( \frac{\omega_0}{\omega_1}\right) ^{-\left( 1+x\right)
/x}\left( \frac{M_s}{\omega_1}\right) ^{-1/x}, 
\label{fincon}
\end{eqnarray}
which, using the following numerical
values for parameters \cite{eu,peak}, 
$\omega_0\sim h\times 10^{-30}\hbox{\rm eV},
\omega_1\sim10^{-1}\hbox{\rm eV},
M_s=g_1\times 10^{28} \hbox{\rm eV} $ (where $0.5\leq h\leq 0.8$ and 
$10^{-3}\laq g_1\laq 10^{-1}$),
 become
\begin{eqnarray}
M&>&h^2g_1\times 10^{10/x-30} \hbox{\rm eV}, \nonumber
\\
z_S&<&h^{-1/x-1}g_1^{-1/x}10^{29-8/x-5/x^2}.
\label{fincon1}
\end{eqnarray}

We now   present a concrete example which serves to demonstrate that a 
reasonable range of parameters exists, in which all conditions are naturally 
satisfied.
We look at axions ($l=1$) with masses below $.1$ eV for a cosmological model
described in \cite{pbb}. In this model 
 $\alpha =-2/\left( d+n+3\right) =-1/6$ and $\beta
=-4d/\left( d+n+3\right) =-1$, for $d=3,n=6$, 
which is in the range specified in Table \ref{tab:SI}.
For this specific model $x=2/3$, 
$ \Omega_{REL}\simeq g_1^2\left( \frac{g_1}{g_S}\right)^2$, 
$\Omega _{NR}\simeq g_1^2\left( \frac{g_1}{g_S}\right)^{2}
\frac {M}{\omega_S}\left( \frac{\omega_m}{\omega_S}\right) ^{-1/3}$.

The ratio of relativistic to nonrelativistic
 energy densities is given by
$
\frac{\Omega _{REL}}{\Omega _{NR}}= \frac{\omega_S}M\left(
\frac{\omega_S}{\omega_m} \right) ^{-1/3},
$
and since $\omega_m=\omega_1\left( M/M_s\right) ^{1/2}$ we obtain
\begin{equation}
\frac{\Omega _{REL}}{\Omega _{NR}}=\frac{\omega_1}{M_s}z_S^{-2/3}\left( 
\frac{M_s}M\right) ^{5/6}.
\label{ratio}
\end{equation}
{}From  conditions (\ref{fincon1}) 
we obtain conditions on $M$ and $z_S$: 
\begin{eqnarray}
M&>&h^2g_1\times 10^{-15}\hbox{\rm eV},
\nonumber \\
1<z_S&<&6 \times 10^{5}\ h^{-5/2}g_1^{-3/2} 
\end{eqnarray}
Taking $10^{-10}\hbox{\rm eV}<M<10^{-2}\hbox{\rm eV},$ for which condition
(\ref{cp3}) is comfortably satisfied, we observe that 
the ratio $\Omega _{REL}:$ $\Omega_{NR}$ can vary in a range from
 well above unity  to well below unity.
For example, choosing $g_1=10^{-1}$ and $g_S=10^{-2}$ 
if the axion's mass is $10^{-10}\hbox{\rm eV}$, and for 
$z_S\sim 2\times 10^{4}$ we
get $\Omega _{REL}:$ $\Omega _{NR}=$ $1:1$ with both energy densities
being near critical, and if we choose
$g_1=10^{-1}$ and $g_S=3\times 10^{-2}$, making $\Omega_{REL}\simeq .1$,
and if $z_S\sim 10^{6}$ we 
obtain
$\Omega _{REL}:$ $\Omega _{NR}=1:10$, with $\Omega _{NR}\simeq 1$, and just 
for fun, the preferred
ratio of 70\% CDM and 20\% HDM \cite{primack} is obtained for the same 
values of couplings, if $z_S\sim 2\times 10^5$. Note that the ratio 
(\ref{ratio}) depends on $z_S^{-2/3} M^{-5/6}$, so the previous examples 
correspond to a range of allowed values.

We now show that it is not possible to obtain
HDM+CDM and produce the CMB fluctuations by the same field, accepting all
conditions on the spectrum. 
If the spectral amplitude is indeed at the level of the observed CMB fluctuations
of about $10^{-5}$, then the spectral slope is also  constrained by the 
data \cite{pdg}.
By parametrizing the slope as $d\Omega /d\ln \omega\sim
\omega^{(n-1)/2}$, we deduce that $n$ can be identified with the  tilt
parameter of the spectrum \cite{seeds1}, which
is  constrained by the data  to be in the range
$
0.8< n< 1.4,
$ which corresponds to 
$
-0.1< x< 0.2.
$
We have required a positive slope, so the available range is just 
$
0< x< 0.2.
$
Now it is a simple exercise to calculate the energy density at $\omega_{m}$
and see that it is much below unity, 
$d\Omega /d\ln \omega(\omega_{m})/d\Omega /d\ln \omega(\omega_{0})=
\left(\frac{\omega_{m}}{\omega_{0}}\right)^x$. Since $\frac{\omega_{m}}{\omega_{0}}
=\frac{\omega_{1}}{\omega_{0}}\left(\frac{M}{M_s}\right)^{1/2}<10^{16}$, and since
$x<0.2$, we can  estimate that this ratio cannot naturally be above $10^{3}$, making
the energy density at $\omega_m$ much less than unity in most of parameter space.
Perhaps, by tuning and forcing parameters into corners, it is possible to find an 
artificial example, but we will not pursue this possibility.

We further show that, in the range of parameters we are
interested in, it is not possible that one scalar field
provides the required CMB fluctuations and a different scalar field
provides HDM+CDM  in a single cosmology.
Since the two fields must be different $l_f\neq l_{HC},$ where the
subscript $f$ denotes a field that is supposed to produce
the CMB anisotropy and the subscript $HC$ denotes the field
that is supposed to produce HDM+CDM. We require
$
\Omega _{REL}^{HC}\simeq g_1^2\left( \frac{g_S}{g_1}\right)^{-2l_{HC}}\simeq 1,
$
and
$
\Omega _{REL}^f\simeq g_1^2\left( \frac{g_S}{g_1}\right)
^{-2l_f}\laq 1.
$
Therefore,
$
\frac{\Omega _{REL}^{HC}}{\Omega _{REL}^f}=\left( \frac{g_S}{g_1}%
\right) ^{2l_f-2l_{HC}}\gaq 1,
$
so for $\frac{g_S}{g_1}<1$ we have to have $l_f<l_{HC}$ and from 
$
x_f=2+2\alpha +l_f\beta 
$ and 
$
x_{HC}=2+2\alpha +l_{HC}\beta 
$
(recall that $\alpha <0$ and $\beta <0$ ) we obtain
$
x_f>x_{HC}.
$
The conclusion is that 
 the only possibility (for  values of $l, -1,0,1)$ is 
$0<x_{HC}<x_f<0.2$. But then $\Omega^{HC}$ dominates  also at the 
lowest frequency, in
contradiction to our assumptions. Therefore, either CMB fluctuations are 
of different origin \cite{seeds1,gv98}, or DM source is different.

If DM in the universe is indeed made of light particles
with gravitational strength interactions its detection in current 
direct searches is extremely difficult, and will probably require new methods
and ideas.

Finally, we note that the models we have described have predictions and 
consequences other than DM. Additional particles such as  
gravitons \cite{bggv} get produced and provide additional possible 
experimental  and observational signatures.

\acknowledgments 

This work is supported in part by the  Israel
Science Foundation administered by the Israel Academy of Sciences and
Humanities.

\end{document}